\documentclass[10pt]{iopart}

\usepackage{iopams}
\usepackage{graphicx}
\usepackage{siunitx}
\usepackage{braket}
\usepackage{cases}
\usepackage[cmyk]{xcolor}
\usepackage{enumerate}
\DeclareSIUnit{\million}{\text{million}}

\begin{document}

\title[Surface engineering for silicon/aluminum resonators]{Substrate surface engineering for high-quality silicon/aluminum superconducting resonators}

\author{C T~Earnest$^{1,2}$, J H~B\'{e}janin$^{1,2}$, T G~McConkey$^{1,3}$, E A~Peters$^{1,2}$, A~Korinek$^{4}$, H~Yuan$^{4}$ and M~Mariantoni$^{1,2}$}

\address{$^{1}$ Institute for Quantum Computing, University of Waterloo, 200 University Avenue West, Waterloo, Ontario N2L 3G1, Canada}
\address{$^{2}$ Department of Physics and Astronomy, University of Waterloo, 200 University Avenue West, Waterloo, Ontario N2L 3G1, Canada}
\address{$^{3}$ Department of Electrical and Computer Engineering, University of Waterloo, 200 University Avenue West, Waterloo, Ontario N2L 3G1, Canada}
\address{$^{4}$ Canadian Centre for Electron Microscopy, Department of Materials Science and Engineering, McMaster University, 1280 Main Street West, Hamilton, Ontario L8S 4M1, Canada}

\ead{matteo.mariantoni@uwaterloo.ca}
\vspace{10pt}

\begin{indented}
\item[] July 20th, 2018
\end{indented}

\begin{abstract}
Quantum bits~(qubits) with long coherence times are an important element for the implementation of medium- and large-scale quantum computers. In the case of superconducting planar qubits, understanding and improving qubits' quality can be achieved by studying superconducting planar resonators. In this Paper, we fabricate and characterize coplanar waveguide resonators made from aluminum thin films deposited on silicon substrates. We perform three different substrate treatments prior to aluminum deposition: One chemical treatment based on a hydrofluoric acid clean; one physical treatment consisting of a thermal annealing at~\SI{880}{\degreeCelsius} in high vacuum; one combined treatment comprising both the chemical and the physical treatments. The aim of these treatments is to remove the two-level state defects hosted by the native oxides residing at the various sample's interfaces. We first characterize the fabricated samples through cross-sectional tunneling electron microscopy acquiring electron energy loss spectroscopy maps of the samples' cross sections. These measurements show that both the chemical and the physical treatments almost entirely remove native silicon oxide from the substrate surface and that their combination results in the cleanest interface. We then study the quality of the resonators by means of microwave measurements in the ``quantum regime'', i.e., at a temperature~$T \sim \SI{10}{\milli\kelvin}$ and at a mean microwave photon number~$\langle n_{\textrm{ph}} \rangle \sim 1$. In this regime, we find that both surface treatments independently improve the resonator's intrinsic quality factor and that the highest quality factor is obtained for the combined treatment, $Q_{\textrm{i}} \sim \SI{0.8}{million}$. Finally, we find that the TLS quality factor averaged over a time period of~\SI{3}{\hour} is~\mbox{$\sim \SI{3}{million}$} at~$\langle n_{\textrm{ph}} \rangle \sim 10$, indicating that substrate surface engineering can potentially reduce the TLS loss below other losses such as quasiparticle and vortex loss.
\end{abstract}

\vspace{1pc}

%\maketitle
\ioptwocol

\section{Introduction}
	\label{sec:Introduction}

Quantum computers capable of outperforming the most advanced classical computers are closer to reality than ever before~\cite{Neill:2018}. Quantum processors comprising a mere~$50$ or $100$ physical quantum bits (qubits) have the potential to show a significant computational advantage over classical supercomputers~\cite{Boixo:2017}. Among the many physical systems used to implement qubits~\cite{Ladd:2010}, superconducting circuits~\cite{Clarke:2008, Wendin:2017} remain one of the most promising candidates to realize medium- and, possibly, large-scale quantum computers~\cite{Gambetta:2017:b}. The past five years have witnessed a steady increase in the size of superconducting quantum computers~\cite{Kelly:2015, Otterbach:2017, Knight:2017, Giles:2018} as well as a continuous increase in qubit lifetime~\cite{Barends:2013, Kamal:2016, Gambetta:2017} and decrease in computational error rates~\cite{Barends:2014, Jeffrey:2014, McKay:2017}.

Scaling up superconducting quantum computers while maintaining low error rates~\cite{Martinis:2015} requires even more accurate qubit control and measurement methods as well as longer qubit lifetimes. Superconducting qubits are made from superconducting thin films (typically aluminum~(Al) or niobium) deposited and patterned above dielectric substrates (typically silicon~(Si) or sapphire). Thus, the two main dissipation phenomena for this type of qubits are associated with quasiparticle loss in the metal~\cite{Serniak:2018} and dielectric loss~\cite{Gambetta:2017:a}. The latter, in particular, is due to native oxide layers that reside at the various qubit interfaces, such as the substrate-metal~(SM) and the substrate-air (vacuum)~(SA) interfaces~\cite{Wisbey:2010, Wenner:2011:b}. Research has shown that these oxide films host so-called two-level state~(TLS) defects~\cite{Anderson:1971, Phillips:1981, Phillips:1987}, which act as a distribution of ``unwanted qubits'' interfering with the actual qubit computational states~\cite{Martinis:2005, Mueller:2017}.

A simple approach to study TLS loss in superconducting devices is realized by fabricating coplanar waveguide~(CPW) resonators and measuring their microwave properties at low temperature, $T \sim \SI{10}{\milli\kelvin}$, and low excitation power or low mean microwave photon number, $\langle n_{\textrm{ph}} \rangle \sim 1$~\cite{Gao:2007}, i.e., at qubit operating conditions. A fundamental characteristic of such resonators is the intrinsic (or internal) quality factor~$Q_{\textrm{i}}$, which allows us to extrapolate the dielectric \textit{loss tangent}~$\tan \delta$~\cite{Collin:2001, McRae:2018}. These measurements make it possible to characterize and improve the device's fabrication process and, thus, to mitigate TLS loss. Additionally, superconducting CPW resonators can be used to investigate and reduce quasiparticle loss~\cite{Nsanzineza:2014}. The ultimate aim of all these studies is to produce superconducting planar resonators with the highest possible~$Q_{\textrm{i}}$ at~$T \sim \SI{10}{\milli\kelvin}$ and $\langle n_{\textrm{ph}} \rangle \sim 1$. While this may not directly translate to correspondingly long-lived qubits (due to the higher complexity of qubit design and fabrication), it provides a reasonable estimate for qubit loss. In addition, implementing very long-lived planar resonators may make it possible to further scale superconducting quantum computing architectures based on so-called ``cat codes''~\cite{Ofek:2016}, which rely heavily on high-quality resonators.

A variety of methods have been shown to reduce TLS loss such as thoughtful material choices~\cite{OConnell:2008, Sage:2011, Martinis:2014}, attention to device geometry and size~\cite{Sage:2011, Gambetta:2017:a}, and fabrication process improvements~\cite{Wisbey:2010, Megrant:2012, Bruno:2015, Richardson:2016, Dunsworth:2017}. Recently, experimental studies have focused on Al films (due to its long-term stability and the self-limiting nature of native Al~oxide) and Si substrates (due to the ease of fabrication and compatibility with classical integrated circuit technology)~\cite{Gambetta:2017:a, Richardson:2016, Dunsworth:2017, Burnett:2018}. At this point in time, the Si/Al planar resonators with the highest quality are those characterized in~\cite{Dunsworth:2017}, with~$Q_{\textrm{i}} \approx \SI{2.5}{\million}$ at low temperature and low mean photon number. The effects of substrate engineering have been well studied and reported for Al films on sapphire substrates~\cite{Megrant:2012, Kamal:2016}; fewer aspects have been investigated in the case of Si/Al resonators~\cite{Wisbey:2010, Dunsworth:2017, Richardson:2016}.

In this Paper, we present a detailed study of substrate surface engineering methods based both on chemical and on physical cleaning treatments. We show the effects of each method on the TLS loss of Si/Al superconducting CPW resonators, using the methods separately and in combination. The compared methods are an \textit{ex situ} ``RCA'' Standard Clean-1~(or \textit{RCA SC-1}) process~\cite{Kern:1993} immediately followed by hydrofluoric~(HF) acid etching (or \textit{HF dip})~\cite{Wisbey:2010}, and two different \textit{in situ} thermal annealing treatments in a high-vacuum~(HV) environment~\cite{Tournet:2016}. We focus on Al films deposited by way of electron-beam evaporation instead of molecular beam epitaxy~(MBE), using the same system we also use to fabricate superconducting qubits; the physical cleaning treatments are carried out in this system.

We first characterize the various methods by means of cross-sectional transmission electron microscope~(TEM) measurements of the SM interface by acquiring electron energy loss spectroscopy~(EELS) maps. We then perform microwave measurements of superconducting CPW resonators and use them to obtain~$Q_{\textrm{i}}$ at low temperature, both at high and at low mean photon number. Additionally, we measure the time fluctuations of the total TLS loss~$F \tan \delta_{\textrm{TLS}}$, where~$F$ is the CPW filling factor~\cite{Collin:2001}, over the course of three-hour long measurements. The resonators fabricated with either the chemical or the physical surface treatment show an improvement over unprocessed devices; a combination of the two treatments results in the biggest improvement with~$Q_{\textrm{i}} \approx \SI{0.8}{\million}$ at~$T \sim \SI{10}{\milli\kelvin}$ and $\langle n_{\textrm{ph}} \rangle \sim 1$ and a time average~$\langle F \tan \delta_{\textrm{TLS}} \rangle \sim \SI{3.5e-07}{}$ or, equivalently, an average TLS quality factor~$\langle Q_{\textrm{TLS}} \rangle \sim \SI{3}{\million}$ at~$T \sim \SI{10}{\milli\kelvin}$ and $\langle n_{\textrm{ph}} \rangle \sim 10$. Aside from characterizing the effects of each individual surface treatment, we demonstrate that a standard electron-beam evaporation system with \textit{in situ} annealing capabilities allows us to reach similar or higher resonator quality compared to an MBE system~\cite{Richardson:2016}.

This Paper is organized as follows. In section~\ref{sec:Methods}, we describe the surface engineering methods used to fabricate the samples studied in this work (see subsection~\ref{subsec:Surface:surface:engineering::Methods}), the microwave measurement setup (see subsection~\ref{subsec:Resonator:measurements::Methods}), and the~TEM and EELS techniques used to perform thin film metrology (see subsection~\ref{subsec:Thin:film:metrology::Methods}). In section~\ref{sec:Results}, we show the thin film metrology (see subsection~\ref{subsec:Thin:film:metrology::Results}) as well as the intrinsic quality factor measurements, including time fluctuation analysis and sample aging effects (see subsection~\ref{subsec:Resonator:measurements::Results}). In section~\ref{sec:Discussion}, we compare our findings to those presented in previous studies and compare the effects of an~\SI{880}{\degreeCelsius} anneal to those of a~\SI{950}{\degreeCelsius} anneal. Finally, in section~\ref{sec:Conclusion}, we summarize our results and outline a procedure that may reduce TLS loss below all other loss mechanisms in planar devices. Additionally, in appendix~\ref{app:Resonator:measurement:and:fitting}, we provide details on the resonator measurement and fitting procedures.

\section{Methods}
	\label{sec:Methods}

The aim of this section is to provide the details required to perform and reproduce the substrate surface engineering methods presented in this Paper. We describe the setup used for the microwave measurements. Finally, we specify the steps followed to prepare the samples for the~TEM and EELS characterization.

\subsection{Substrate surface engineering: Methods}
	\label{subsec:Surface:surface:engineering::Methods}

Five samples are fabricated in order to study the effects of various substrate surface engineering methods: One control sample with no substrate surface treatment (``Unprocessed''); one sample prepared with a chemical cleaning treatment that comprises an RCA SC-1 process and an HF dip (``RCA 1 + HF''); one sample prepared with a physical treatment consisting of a short thermal annealing step at~\SI{880}{\degreeCelsius} (``\SI{880}{\degreeCelsius} Anneal''); one sample prepared with a long thermal annealing step at~\SI{950}{\degreeCelsius} (``\SI{950}{\degreeCelsius} Anneal''); one sample prepared with both the chemical cleaning treatment and the short thermal annealing step (``RCA 1 + HF + \SI{880}{\degreeCelsius} Anneal''). These treatments are all applied to the substrate directly before the deposition of the Al thin film.

All the samples for this study are fabricated on high-resistivity
($> \SI{10}{\kilo\ohm\centi\meter}$) \SI{500}{\micro\meter} thick 4-in.~float-zone~(FZ) undoped Si~$(100)$ wafers. The details of the surface engineering methods are:
\begin{enumerate}
\item \textit{RCA 1 + HF.}\textemdash A~\SI{10}{\minute} RCA SC-1 process (a bath of~$NH_4OH \colon H_2O_2 \colon H_2O, 1 \colon 1 \colon 5, \textrm{at } \SI{75}{\degreeCelsius}$), immediately followed by a~\SI{1}{\minute} bath in buffered oxide etchant containing~\SI{1}{\percent} HF acid. After cleaning, the wafers are dried with nitrogen gas and placed immediately (within~\mbox{$\sim \SI{10}{\minute}$}) in the load-lock of the deposition system;

\item \textit{\SI{880}{\degreeCelsius} Anneal.}\textemdash A~\SI{30}{\minute} ramp to~\SI{880}{\degreeCelsius}, followed by a~\SI{10}{\minute} anneal at~\SI{880}{\degreeCelsius} in~HV at a pressure~\mbox{$\lesssim \SI{4e-7}{\milli\bar}$};

\item \textit{\SI{950}{\degreeCelsius} Anneal.}\textemdash A~\SI{30}{\minute} ramp to~\SI{950}{\degreeCelsius}, followed by a~\SI{60}{\minute} anneal at~\SI{950}{\degreeCelsius} in~HV at a pressure~\mbox{$\lesssim \SI{4e-7}{\milli\bar}$}. This is the highest temperature attainable in our system.
\end{enumerate}
At the end of each annealing process, the sample is allowed to cool overnight in the same chamber down to a temperature of~\mbox{$\approx \SI{24}{\degreeCelsius}$} before \textit{in situ} Al deposition. During the cooldown period, the chamber reaches ultra-high vacuum~(UHV) with a pressure~\mbox{$\sim \SI{1e-10}{\milli\bar}$}. The unprocessed sample is a brand-new wafer that is not submitted to any chemical or physical treatment prior to Al deposition.

The annealing temperatures are measured by means of a type~K thermocouple in the heating element, which is calibrated to the temperature of the wafer during commissioning via a thermocouple attached to the front side of the wafer (the heating element heats from the back side). We estimate that the actual temperature of the wafer can fluctuate by~\mbox{$\sim \mp \SI{15}{\degreeCelsius}$}; this uncertainty is obtained from a set of tests with a thermocouple on the front of a molybdenum plate. We notice that the center of a 4-in.~wafer is more homogeneously heated than the wafer edges; thus, we characterize dies only from the wafer center.

The~Al films are deposited by means of an~UHV electron-beam
physical vapor-deposition~(EBPVD) system from Plassys-Bestek SAS, model MEB 550 SL3-UHV. Each film is~\SI{150}{\nano\meter} thick and is deposited at a rate of~\SI{2}{\angstrom\per\second} from a~\SI{99.999}{\percent} pure Al shot with a large pellet size to reduce the surface-to-volume ratio. The base pressure in the EBPVD chamber is~\mbox{$\sim \SI{e-10}{\milli\bar}$} before evaporation and \mbox{$\sim \SI{2.0e-8}{\milli\bar}$} during evaporation.

Each sample is patterned by means of optical lithography using a~\SI{1}{\micro\meter}-thick layer of Microposit S$1811$ positive photoresist from the Shipley Company-MicroChem Corp. The photoresist is developed in Microposit MF-$319$ developer also from Shipley and then etched using an inductively coupled plasma~(ICP) etcher from Oxford Instruments plc, model Plasmalab System~$100$ ICP$380$ III-V and metals chlorine etcher. Each sample is immediately transferred into a water bath after etching to remove any remaining chlorinated species on the sample surface and kept soaked in water for at least~\SI{10}{\minute}. The remaining photoresist is removed in a heated solvent bath. Additional photoresist is applied as a protective layer before dicing and again removed in heated solvents after dicing.

\subsection{Resonator measurements: Methods}
	\label{subsec:Resonator:measurements::Methods}

The diced samples are mounted in a quantum socket package analogous to that described in~\cite{Bejanin:2016} and anchored to the mixing chamber of a dilution refrigerator with base temperature~$T \sim \SI{10}{\milli\kelvin}$. The measurement setup, which features heavily attenuated microwave lines as well as careful infrared and magnetic shielding, is described in detail in~\cite{Bejanin:2016}. The total attenuation of the input microwave channel that connects the room-temperature electronics to the samples is~\mbox{$\sim \SI{76}{\decibel}$} at~\SI{5}{\giga\hertz}. All measurements are performed by means of a vector network analyzer~(VNA) from Keysight Technologies Inc., model~PNA-L N5230A.

\subsection{Thin film metrology: Methods}
	\label{subsec:Thin:film:metrology::Methods}

The material effects of each substrate surface treatment are studied through cross-sectional~TEM measurements. Cross sections of the samples are prepared within two different types of focused ion beam~(FIB) systems; one system from Carl Zeiss AG, model NVision~40 and a brand-new system from Thermo Fisher Scientific, model Helios~G4 plasma~FIB (or pFIB). The NVision~40 FIB system works on traditional gallium+~(Ga+) ions to mill the samples, whereas the pFIB system works on xenon+~(Xe+) ions. Xenon+ ions do not react with the oxygen~(O) content of the samples and, thus, do not induce additional sample contamination~\cite{Mayer:2007}. In fact, we use Ga+ ions only for the~RCA 1 + HF and RCA 1 + HF + \SI{880}{\degreeCelsius} Anneal samples, where the~O content is small. The~Unprocessed and \SI{880}{\degreeCelsius} Anneal samples are first milled with~Ga+ ions; after noticing considerable contamination, we milled new samples using~Xe+ ions with the newly acquired~pFIB system. The other two samples show very small~Ga+ ion implantation and, thus, are not milled using the~pFIB system.

After depositing a mixture of carbon and platinum~(Pt) on the sample's surface to prevent damage, the FIB systems are used to prepare thin sample's sections. Then, two trenches are milled on both sides of the area of interest using either a~Ga+ or Xe+ ion beam. The resulting thin section is lifted out from the sample using a micromanipulator needle. The cross section is subsequently attached to a copper~TEM grid using~Pt. Finally, the cross section is thinned to electron transparency, \mbox{$\sim \SI{80}{\nano\meter}$} thickness, followed by a low-voltage cleaning at~\SI{5}{\kilo\volt} for the~Ga+ ion beam and \SI{2}{\kilo\volt} for the~Xe+ ion beam.

The cross sections are imaged in a~TEM system from the FEI Company, model Titan~$80-300$, operated at~\SI{200}{\kilo\volt}. The system is equipped with a~CEOS image and probe corrector and a Gatan imaging filter from Gatan, Inc. (type ``Quantum Energy Filter''). Low magnification scanning~TEM images are acquired to assess the overall quality of the sample and the roughness of the Si/Al interface. High-angle annular dark-field~(HAADF) high-resolution scanning~TEM micrographs are acquired to visualize the interface between the~Si substrate and the~Al film.

EELS maps are acquired to investigate the formation of an oxide layer at the Si/Al interface. The maps are acquired with a step size of~\SI{2}{\angstrom}; the dwell time per spectrum is set to~\SI{0.01}{\second}, with a detector binning of~$[1, 130]$ and \SI{1}{\electronvolt} per channel. The convergence semi-angle of the electron beam is set to~\SI{19.1}{\milli\radian} and the collection semi-angle to~\SI{55}{\milli\radian}. The beam current is set to~\mbox{$\approx \SI{150}{\pico\ampere}$} in order to avoid beam induced sample damage.

EELS elemental quantification is calculated using the Gatan digital micrograph~$3.22$ software. The spectra are aligned to eliminate energy shifts during acquisition. For the background fit of the edges a power law model is fitted to the spectra. The Hartree-Slater cross-section model is used for quantification. The energy loss near edge structure~(ELNES) region of the peaks is excluded up to~\SI{40}{\electronvolt}.

\begin{figure*}[t!]
	\centering
	\includegraphics{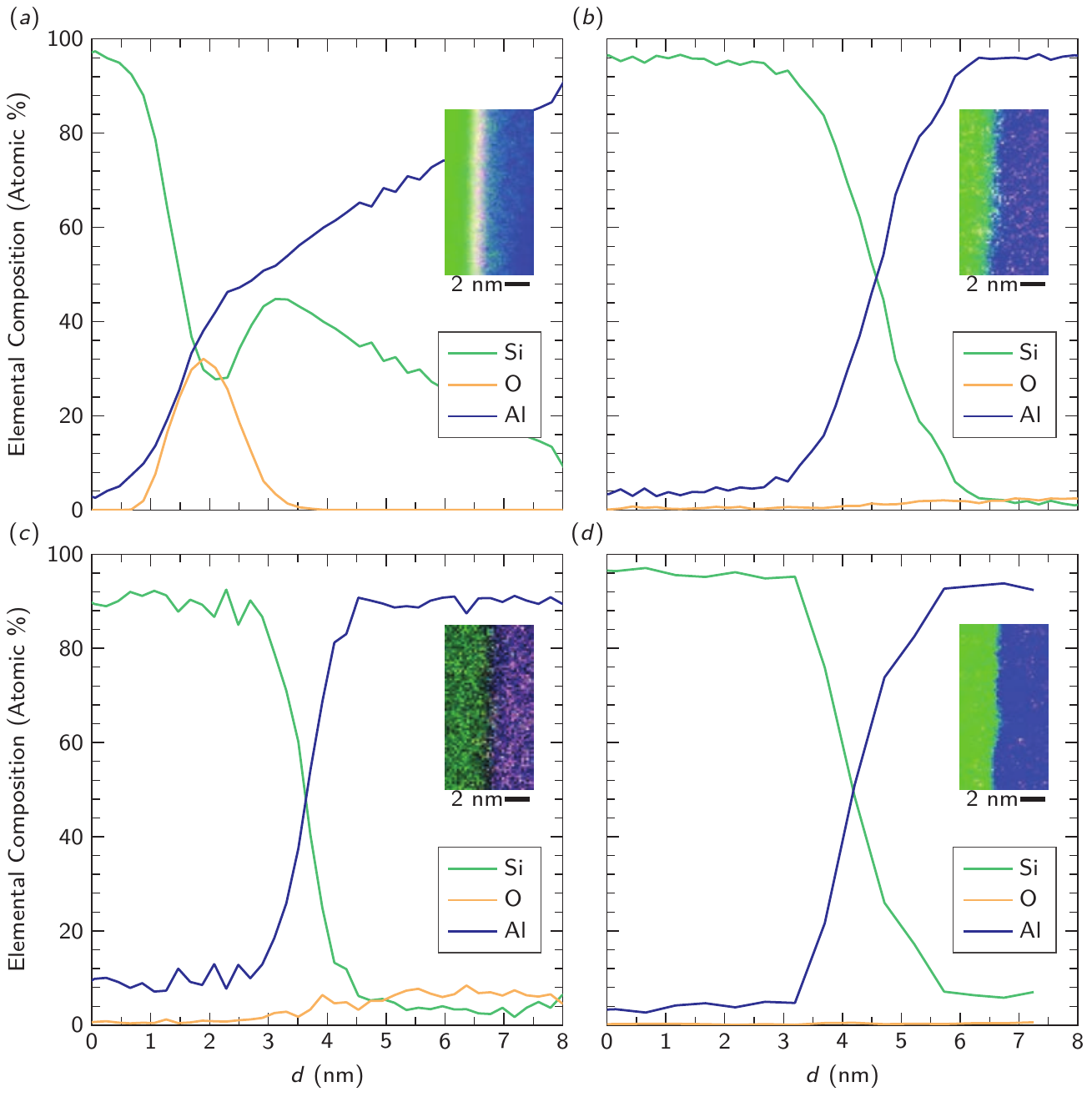}
\caption{Elemental composition as atomic percentage vs.~$d$ for~Si, O, and Al at the SM interface of each sample. Data is collected with~EELS over a rectangular area~(inset) and averaged to give the atomic percentage shown in the line graph. Si: Green (medium gray); O: Light orange (light gray); Al: Blue (dark gray). (a) Unprocessed sample (pFIB). (b) RCA 1 + HF sample (Ga+~FIB). (c) \SI{880}{\degreeCelsius} Anneal (pFIB). (d) RCA 1 + HF + \SI{880}{\degreeCelsius} Anneal sample (Ga+~FIB).}
	\label{figure1a_1d_earnest}
\end{figure*}

\begin{figure*}[t!]
	\centering
	\includegraphics{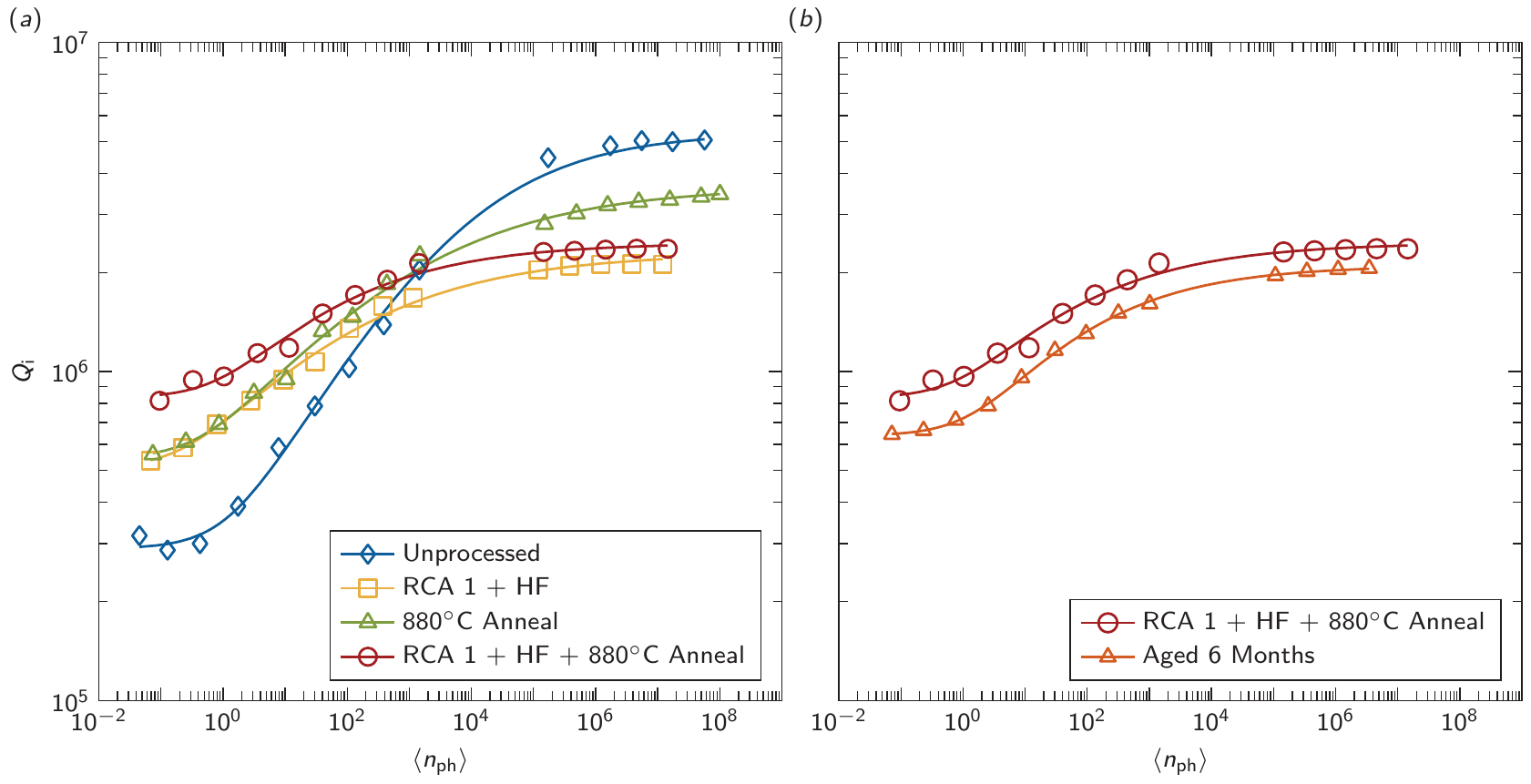}
\caption{Resonators' S-curve measurements. Symbols correspond to data and solid lines to fitting curves obtained using the modified photon-dependent TLS model of~(\ref{eq:TLS:fit}). All measurements are performed at~$T \sim \SI{10}{\milli\kelvin}$ (see main text for details). (a) Intrinsic quality factor~$Q_{\textrm{i}}$ vs.~mean photon number~$\langle n_{\textrm{ph}} \rangle$ for one resonator on each of the four samples: Uprocessed, RCA 1 + HF, \SI{880}{\degreeCelsius} Anneal, and RCA 1 + HF + \SI{880}{\degreeCelsius} Anneal. (b) The data for the resonator on sample~RCA 1 + HF + \SI{880}{\degreeCelsius} Anneal in~(a) are compared to those for a resonator fabricated on a similar sample, but aged over a time period of~\SI{6}{months}.}
	\label{figure2a_2b_earnest}
\end{figure*}

\section{Results}
	\label{sec:Results}

In this section, we show the efficacy of the various substrate surface engineering methods by presenting both thin film metrology and microwave measurements. In particular, we study quality factor time fluctuations and sample aging.

\subsection{Thin film metrology: Results}
	\label{subsec:Thin:film:metrology::Results}

Figure~\ref{figure1a_1d_earnest} shows the elemental composition as atomic percentage for~Si, O, and Al at the SM interface of each sample. As expected, these measurements demonstrate that the~RCA 1 + HF chemical cleaning treatment is effective at removing almost entirely the~Si native oxide and that this oxide does not reform in the time between cleaning and placement into the~Al deposition system. In addition, the measurements show that the~\SI{880}{\degreeCelsius} Anneal also removes almost entirely the~Si native oxide present on the substrate. When applied separately, both the chemical and physical cleaning treatments result in noticeable~O implantation in the~Al region. The combined treatment, instead, not only cleans the interface from any oxide, but also leads to a very sharp transition between the~Si and Al regions, without any~O interdiffusion or implantation. In section~\ref{sec:Discussion}, we compare these results with those obtained using the~\SI{950}{\degreeCelsius} Anneal.

\subsection{Resonator measurements: Results}
	\label{subsec:Resonator:measurements::Results}

\begin{figure}[b!]
	\centering
	\includegraphics{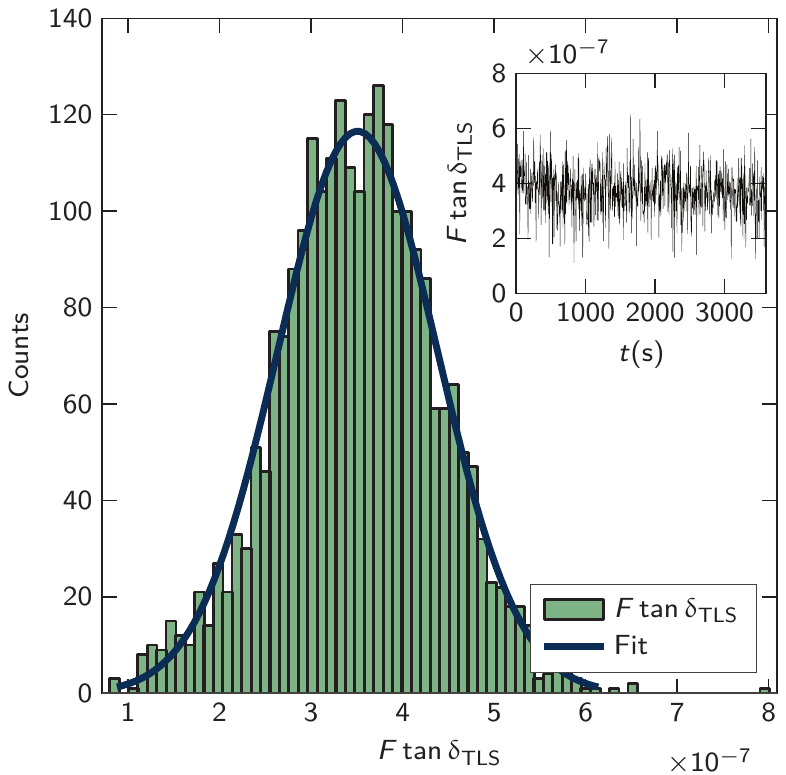}
\caption{Time fluctuations of TLS loss. TLS loss~$F \tan \delta_{\textrm{TLS}}$ vs.~time~$t$ plotted as a histogram comprising~$70$ bins. The measurement is performed over a time period of~\SI{3}{\hour} (see main text for details). The solid line is a fit to a normal distribution. Inset: Time trace associated with the histogram, showing~$F \tan \delta_{\textrm{TLS}}$ vs.~$t$ up to~$t = \SI{3600}{\second}$.}
	\label{figure3_earnest}
\end{figure}

Each sample comprises a feed CPW transmission line capacitively coupled with coupling strength~$\kappa$ to nine meandered quarter-wave resonators each with a different resonance frequency~$\tilde{f}_0$, in a multiplexed design (see the works in~\cite{Bejanin:2016, McRae:2017} for similar designs). We measure the feed-line transmission coefficient~$S_{21}$ in the frequency range between~$4$ and \SI{8}{\giga\hertz} and, for consistency, select one resonator from each sample with the same designed~$\tilde{f}_0 = \SI{4.5}{\giga\hertz}$ and $\kappa = \SI{40}{\kilo\hertz}$. The resonator transmission measurements are fitted as described in appendix~\ref{app:Resonator:measurement:and:fitting} to give the actual resonance frequency~$f_0$, rescaled coupling quality factor~$Q^{\ast}_{\textrm{c}}$~\cite{Megrant:2012}, and internal quality factor~$Q_{\textrm{i}}$ at the highest and lowest measured mean photon number, $Q_{\textrm{HP}}$ and $Q_{\textrm{LP}}$, respectively (see table~\ref{table:resonator:parameters}). The mean photon number~$\langle n_{\textrm{ph}} \rangle$ is estimated from the input excitation power (see appendix~\ref{app:Resonator:measurement:procedure}). All measurements are performed at~$T \sim \SI{10}{\milli\kelvin}$.

Figure~\ref{figure2a_2b_earnest}~(a) shows~$Q_{\textrm{i}}$ as a function of~$\langle n_{\textrm{ph}} \rangle$, with characteristic~\textit{S-curve} shape~\cite{McRae:2018}, for four resonators on samples fabricated using different substrate surface treatments. As expected, the highest~$Q_{\textrm{LP}}$ is found for the resonator on the~RCA 1 + HF + \SI{880}{\degreeCelsius} Anneal sample, $Q_{\textrm{LP}} \approx \SI{0.82e+06}{}$. Notably, the resonators on the~RCA 1 + HF and the~\SI{880}{\degreeCelsius} Anneal samples have both~$Q_{\textrm{LP}} \approx \SI{0.55e+06}{}$, indicating that the separate chemical and physical substrate surface treatments lead to similar improvements compared to the resonator on the~Unprocessed sample, which has~$Q_{\textrm{LP}} \approx \SI{0.32e+06}{}$.

In order to obtain a more accurate estimate of the resonator TLS loss, we invert the data of each S-curve and fit it to the modified photon-dependent TLS model~\cite{Richardson:2016, Burnett:2018}
\begin{equation}
\frac{1}{Q_{\textrm{i}} \left( \langle n_{\textrm{ph}} \rangle \right)} \simeq F \tan \delta^0_{\textrm{TLS}} \left(1 + \frac{\langle n_{\textrm{ph}} \rangle}{\langle n^{}_{\textrm{ph}} \rangle^{\textrm{c}}}\right)^{- \alpha} + \frac{1}{Q^{\ast}} \, ,
	\label{eq:TLS:fit}
\end{equation}
where~$F \tan \delta^0_{\textrm{TLS}}$ is the total TLS loss at zero photon number and zero temperature, $\langle n^{}_{\textrm{ph}} \rangle^{\textrm{c}}$ is a critical mean photon number above which the TLS defects start to be saturated, $\alpha$ is an exponent indicating the deviation from the standard TLS model (in the standard model~$\alpha = 1 / 2$), and the constant offset~$1 / Q^{\ast}$ accounts for all other (non-TLS) losses ($1 / Q^{\ast} \sim 1 / Q_{\textrm{HP}}$). The fitting curves are overlaid to the data in figure~\ref{figure2a_2b_earnest}. The fitted values of~$\langle n^{}_{\textrm{ph}} \rangle^{\textrm{c}}$, $\alpha$, $F \tan \delta^0_{\textrm{TLS}}$, and $1 / Q^{\ast}$, as well as~\cite{Calusine:2018}
\begin{equation}
F \tan \delta_{\textrm{TLS}} \simeq \frac{1}{Q_{\textrm{LP}}} - \frac{1}{Q_{\textrm{HP}}}
	\label{eq:F:tan:delta:TLS}
\end{equation}
are reported in table~\ref{table:resonator:parameters}.

\begin{table*}[t!]
\caption{Quantitative analysis of loss mechanisms for one resonator on each of four samples prepared with different substrate surface treatments. $f_0$: Fitted resonance frequency; $Q^{\ast}_{\textrm{c}}$: Fitted rescaled coupling quality factor; $Q_{\textrm{LP}}$: Fitted internal quality factor at the lowest measured mean photon number; $\langle n^{}_{\textrm{ph}} \rangle^{\textrm{c}}$: Critical mean photon number (see main text for details); $\alpha$: Deviation from the standard TLS model; $F \tan \delta^0_{\textrm{TLS}}$: Total TLS loss at zero photon number and zero temperature fitted from~(\ref{eq:TLS:fit}); $F \tan \delta_{\textrm{TLS}}$: Approximate total TLS loss estimated as in~(\ref{eq:F:tan:delta:TLS}); $1 / Q^{\ast}$: Non-TLS loss fitted from~(\ref{eq:TLS:fit}); $1 / Q_{\textrm{HP}}$: Approximate non-TLS loss.}
	\label{table:resonator:parameters}
\footnotesize
%\centering
%\begin{indented}
\lineup
\begin{tabular*}{\textwidth}{@{{ }}lc*{9}{@{\extracolsep{0pt plus
6pt}}c}}
\br
	Sample & $f_0$ & $Q^{\ast}_{\textrm{c}}$ & $Q_{\textrm{LP}}$ & $\langle n^{}_{\textrm{ph}} \rangle^{\textrm{c}}$ & $\alpha$ & $F \tan \delta^0_{\textrm{TLS}}$ & $F \tan \delta_{\textrm{TLS}}$ & $1 / Q^{\ast}$ & $1 / Q_{\textrm{HP}}$ \cr
	       & \raisebox{-3pt}{(\SI{}{\giga\hertz})} & \raisebox{-3pt}{\SI{e+06}{}} & \raisebox{-3pt}{\SI{e+06}{}} & \raisebox{-3pt}{\m---} & \raisebox{-3pt}{---} & \raisebox{-3pt}{\SI{e-06}{}} & \raisebox{-3pt}{\SI{e-06}{}} & \raisebox{-3pt}{\SI{e-06}{}} & \raisebox{-3pt}{\SI{e-06}{}} \cr
\mr
	Unprocessed & $4.487$ & $0.39$ & $0.32$ & $1.17$ & $0.33$ & $3.27$ & $2.93$ & $0.19$ & $0.20$ \cr
	RCA 1 + HF & $4.501$ & $0.27$ & $0.53$ & $0.21$ & $0.25$ & $1.53$ & $1.42$ & $0.44$ & $0.47$ \cr
	\SI{880}{\degreeCelsius} Anneal & $4.512$ & $0.34$ & $0.56$ & $0.40$ & $0.24$ & $1.56$ & $1.50$ & $0.27$ & $0.29$ \cr
	RCA 1 + HF + \SI{880}{\degreeCelsius} Anneal & $4.506$ & $0.34$ & $0.82$ & $0.78$ & $0.28$ & $0.80$ & $0.80$ & $0.41$ & $0.42$ \cr
\br
\end{tabular*}
%\end{indented}
\end{table*}
\normalsize

Figure~\ref{figure2a_2b_earnest}~(b) shows~$Q_{\textrm{i}} ( \langle n_{\textrm{ph}} \rangle )$ for one resonator on an~RCA 1 + HF + \SI{880}{\degreeCelsius} Anneal sample fabricated at the same time as previous samples, but aged over a time period of~\mbox{$\sim \SI{6}{months}$}. The sample is stored in atmosphere in a cleanroom environment prior to measurement. Aging effects on the resonator quality are very small and uniform as a function of~$\langle n_{\textrm{ph}} \rangle$.

Finally, we examine the fluctuations of the TLS loss~$F \tan \delta_{\textrm{TLS}}$ as a function of time~$t$ by repeating the same transmission measurement for the~RCA 1 + HF + \SI{880}{\degreeCelsius} Anneal sample at~$\langle n_{\textrm{ph}} \rangle \approx 10$ over a period of~\SI{3}{\hour}. Each measurement lasts~\mbox{$\approx \SI{3}{\second}$} and the measurements are carried out in succession without a predetermined timing. Each transmission measurement is fitted and~$F \tan \delta_{\textrm{TLS}}$ is found as in~(\ref{eq:F:tan:delta:TLS}). Figure~\ref{figure3_earnest} shows~$F \tan \delta_{\textrm{TLS}} ( t )$ as a histogram, with an inset displaying the raw time trace truncated at~$t = \SI{3600}{\second}$. The histogram mean value (time average) is~$\langle F \tan \delta_{\textrm{TLS}} \rangle \sim \SI{3.5e-07}{}$, with a standard deviation~\mbox{$\sim \SI{0.9e-07}{}$}.

\section{Discussion}
	\label{sec:Discussion}

It is worth comparing our results with those reported in other studies. In~\cite{Megrant:2012}, it was found that cleaning a sapphire substrate by means of activated molecular oxygen~O$^{\ast}_2$ while heating the substrate in~HV at~\SI{850}{\degreeCelsius} or only heating the substrate in~UHV at~\SI{850}{\degreeCelsius} has the most significant benefits on the CPW resonator quality in the quantum regime. We find similar results by chemically cleaning a~Si substrate \textit{ex situ} and then annealing it at~\SI{880}{\degreeCelsius} in~HV. Additionally, our results are based on electron-beam evaporated~Al instead of MBE~Al. In fact, our highest~$Q_{\textrm{LP}}$ is slightly higher than the highest~$Q_{\textrm{LP}}$ obtained in~\cite{Megrant:2012} using electron-beam evaporation without any special clean.

Compared to the study in~\cite{Wisbey:2010}, we add a thermal annealing step to the overall cleaning process. This step increases~$Q_{\textrm{LP}}$ by~\mbox{$\sim \SI{50}{\percent}$} when combined with an \textit{ex situ} chemical clean. In~\cite{Dunsworth:2017}, it is shown that by performing an~HF dip clean of a~Si substrate followed by annealing at~\SI{900}{\degreeCelsius} in an~UHV MBE system results in a~$(2 \times 1)$ surface reconstruction. This is a very similar surface treatment as the combined clean described in our work. The resonators fabricated on such a substrate using an~Al electron-beam evaporator attached to the MBE annealing chamber have~$Q_{\textrm{LP}} \sim \SI{2}{\million}$, which is approximately a factor of~$2$ larger than for our best resonator. A similar result was found in~\cite{Richardson:2016}, where, however, the~Al film were grown using an~MBE system and the best quality factor is slightly lower than ours. An important addition of our work is to clearly break down the cleaning process into separate steps, characterizing each step using both thin film metrology and microwave characterization.

It is worth noting that the quality factors at high mean photon number, $Q_{\textrm{HP}}$, shown in figure~\ref{figure2a_2b_earnest}~(a) vary significantly for the different S-curves, which, in fact, cross each other when increasing~$\langle n_{\textrm{ph}} \rangle$. A similar phenomenon was observed in~\cite{Dunsworth:2017}. This means that while the TLS loss clearly decreases when applying suitable surface treatments, the non-TLS loss does not. Further investigations will be required to understand this issue, which may become important when the TLS loss will eventually be totally removed from the samples and non-TLS losses will be the main dissipation channels.

\begin{figure*}[t!]
	\centering
	\includegraphics{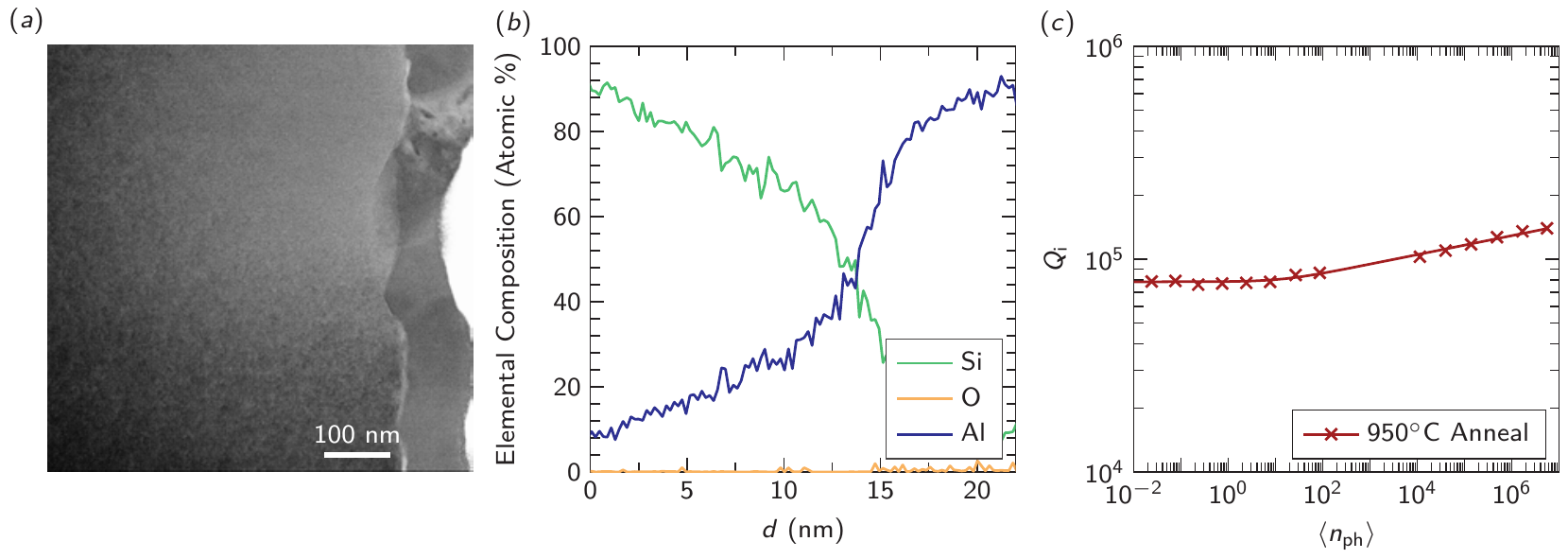}
\caption{\SI{950}{\degreeCelsius} Anneal sample. (a) Cross-sectional~TEM micrograph of the sample. (b) Atomic percentage of Si, O, and Al at the sample's SM interface. Si: Green (medium gray); O: Light orange (light gray); Al: Blue (dark gray). Sample preparation perfromed using pFIB. Data is collected with~EELS, as in figure~\ref{figure1a_1d_earnest}. (c) $Q_{\textrm{i}}$ vs.~$\langle n_{\textrm{ph}} \rangle$ (crosses) with fitting curve obtained from~(\ref{eq:TLS:fit}) (solid line). In this case, $\alpha \sim 0.05$, indicating a large departure from the stanadrd TLS model.}
	\label{figure4a_4c_earnest}
\end{figure*}

In addition to the~\SI{10}{\minute} anneal at~\SI{880}{\degreeCelsius}, we perform a more aggressive~\SI{950}{\degreeCelsius} anneal for~\SI{60}{\minute}. As elucidated by figure~\ref{figure4a_4c_earnest}~(a) and (b), this treatment completely removes the native Si~oxide at the SM interface. However, it causes significant surface roughness and damage to the crystal structure at the Si~substrate surface. These effects outweigh any benefit from the oxide removal, yielding a resonator with very low~$Q_{\textrm{i}}$ for all values of~$\langle n_{\textrm{ph}} \rangle$ (see figure~\ref{figure4a_4c_earnest}~(c)). This result implies that annealing time and temperature should be limited to the minimum required to remove any remaining oxide layer after the chemical clean.

\section{Conclusion}
	\label{sec:Conclusion}

We show that a chemical (HF dip) or physical (thermal annealing at~\SI{880}{\degreeCelsius} in~HV) treatment of a Si~substrate surface positively impact the microwave performance of~Al superconducting CPW resonators fabricated on such substrates. The~Al films are deposited by means of electron-beam evaporation in an~UHV environment, with a deposition system commonly used to fabricate superconducting qubits. No~Si trenching is performed on any of our devices. The highest internal quality factor at low temperature ($T \sim \SI{10}{\milli\kelvin}$) and low mean microwave photon number ($\langle n_{\textrm{ph}} \rangle \sim 1$) is obtained by performing both substrate treatments in succession, $Q_{\textrm{LP}} \sim \SI{0.8}{\million}$. Aging effects on~$Q_{\textrm{LP}}$ are tested by measuring a resonator after storing the sample in a cleanroom environment for~\mbox{$\sim \SI{6}{months}$}. We find that aging has very small and uniform effects on the resonator quality.

Time fluctuations of the TLS loss for the resonator with the highest quality factor are estimated by measuring the loss over a time period of~\SI{3}{\hour}. We find a time-average TLS quality factor~$\langle Q_{\textrm{TLS}} \rangle \sim \SI{3}{\million}$ at~$T \sim \SI{10}{\milli\kelvin}$ and $\langle n_{\textrm{ph}} \rangle \sim 10$, which is similar to the highest TLS quality factor measured in~\cite{Calusine:2018}. In that work, such a high quality factor is obtained when trenching the~Si gap of the CPW line by~\SI{2.2}{\micro\meter}, i.e., a CPW resonator with a deep trench. This result suggests that by trenching our~Si/Al resonators it may be possible to further enhance~$\langle Q_{\textrm{TLS}} \rangle$ by a factor of~$2$. This would result in~$\langle Q_{\textrm{TLS}} \rangle \sim \SI{6}{\million}$, which is significantly larger than the typical internal quality factor at high mean microwave photon number of our resonators, $Q_{\textrm{HP}} \sim \SI{2.5}{\million}$. Under these conditions, TLS defects would not be the limiting channel for dissipation, which, instead, would likely be dominated by quasiparticles, vortices, metal surface roughness, and radiative loss channels.

We characterize the film properties focusing on the SM interface by means of cross-sectional~TEM measurements. These measurements show that both the~HF dip chemical cleaning and thermal annealing at~\SI{880}{\degreeCelsius} remove almost entirely the native Si~oxide layer at the SM interface. In both cases, we observe a small, but noticeable~O implantation in the~Al region. Such implantation is absent when applying a combined chemical and physical treatment, which results in the highest~$Q_{\textrm{LP}}$. In addition, we find that a more aggressive and longer thermal anneal at~\SI{950}{\degreeCelsius} damages and roughens the substrate surface, negating any positive impact on the resonator quality.

\section*{Acknowledgements}

This research was undertaken thanks in part to funding from the Canada First Research Excellence Fund~(CFREF) and the Discovery and Research Tools and Instruments Grant Programs of the Natural Sciences and Engineering Research Council of Canada~(NSERC). Electron microscopy was performed at
the Canadian Centre for Electron Microscopy (also supported
by NSERC and other government agencies). We would like to acknowledge the Canadian Microelectronics Corporation~(CMC) Microsystems for the provision of products and services that facilitated this research, as well as the Quantum NanoFab Facility at the University of Waterloo and Nathan Nelson-Fitzpatrick for his help with the~EBPVD system from Plassys-Bestek. MM acknowledges his fruitful discussions with Zbigniew R Wasilewski.

\clearpage
\newpage

\appendix
\section*{Appendix}
\setcounter{section}{1}
	\label{app:Resonator:measurement:and:fitting}

In this appendix, we provide details on the resonator measurement and fitting procedures used to find the results presented in the main text. We use a systematic experimental routine that allows us to obtain reliable and repeatable measurements of the intrinsic quality factor of resonators, $Q_{\textrm{i}}$, as a function of the excitation power. This power corresponds to an equivalent mean microwave photon number, $\langle n_{\textrm{ph}} \rangle$. Our routine is initialized with the approximate resonance frequency~$f_0$ and frequency span~$\Delta f = f - f_0$ of the resonator to be measured; from these initial values, the routine automatically finds proper measurement ranges and normalization traces, as well as the number of traces to be averaged in order to obtain a fitting uncertainty below a certain threshold. The power is then decreased and the measurement repeated with updated parameters until the minimum desired power level is reached.

We now explain the routine in more detail. We separate the resonator measurement procedure from the resonator fitting procedure, since they are independent.

\subsection{Resonator fitting procedure}
	\label{app:Resonator:fitting:procedure}

\begin{figure}[t!]
	\centering
	\includegraphics{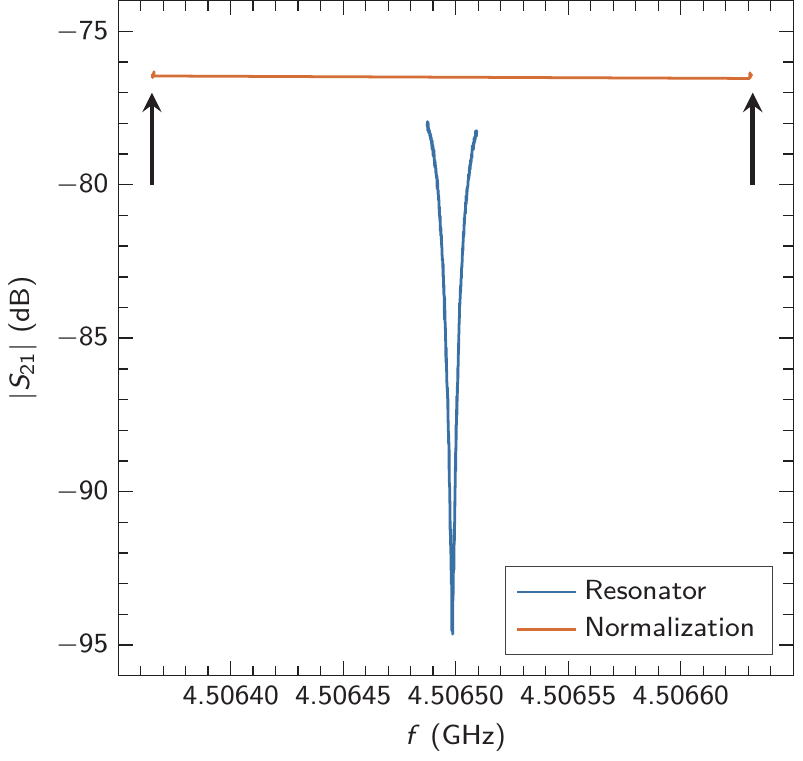}
\caption{Normalization and resonator data. The resonator data is localized in proximity of~$f_0$ (``Resonator''; blue (dark gray)). The normalization data is a set of~$8$ data points in two regions indicated by arrows. These two regions are fitted with a line (``Normalization''; orange (light gray)).}
	\label{figure5_earnest}
\end{figure}

\begin{figure*}[t!]
	\centering
	\includegraphics{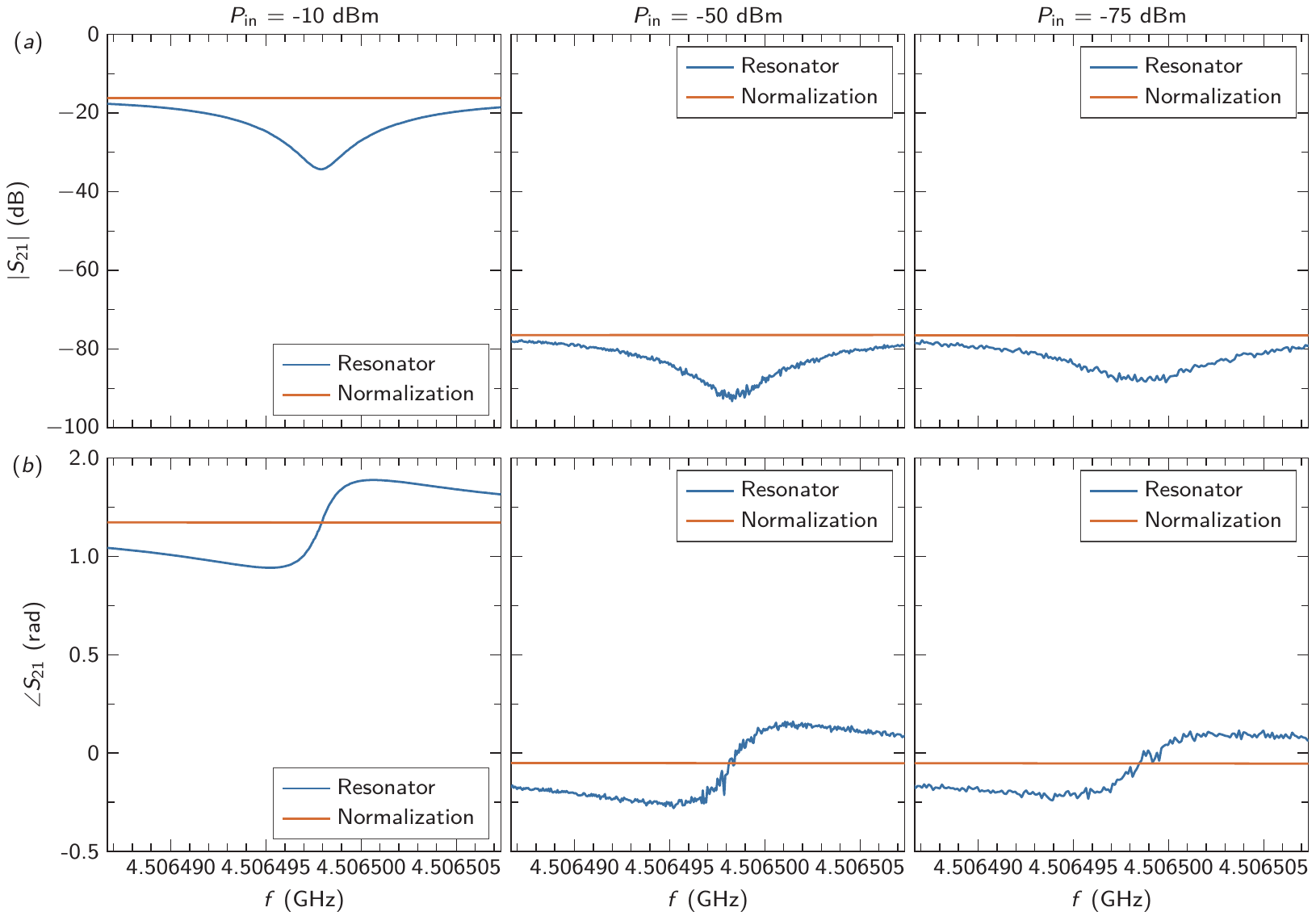}
\caption{Resonator measurement procedure. Each panel comprises three columns, one for each different excitation power (see appendix text for details). (a) $| S_{21} |$ vs.~$f$; resonator data (``Resonator''; blue (dark gray)) and normalization line (``Normalization''; orange (light gray)). (b) Same as in~(a), but for~$\angle S_{21}$ vs.~$f$. In both panels, the normalization lines originate from a set of~$8$ data points in two regions far on the left and right of resonance (see figure~\ref{figure5_earnest}).}
	\label{figure6a_6b_earnest}
\end{figure*}

\begin{figure*}[t!]
	\centering
	\includegraphics{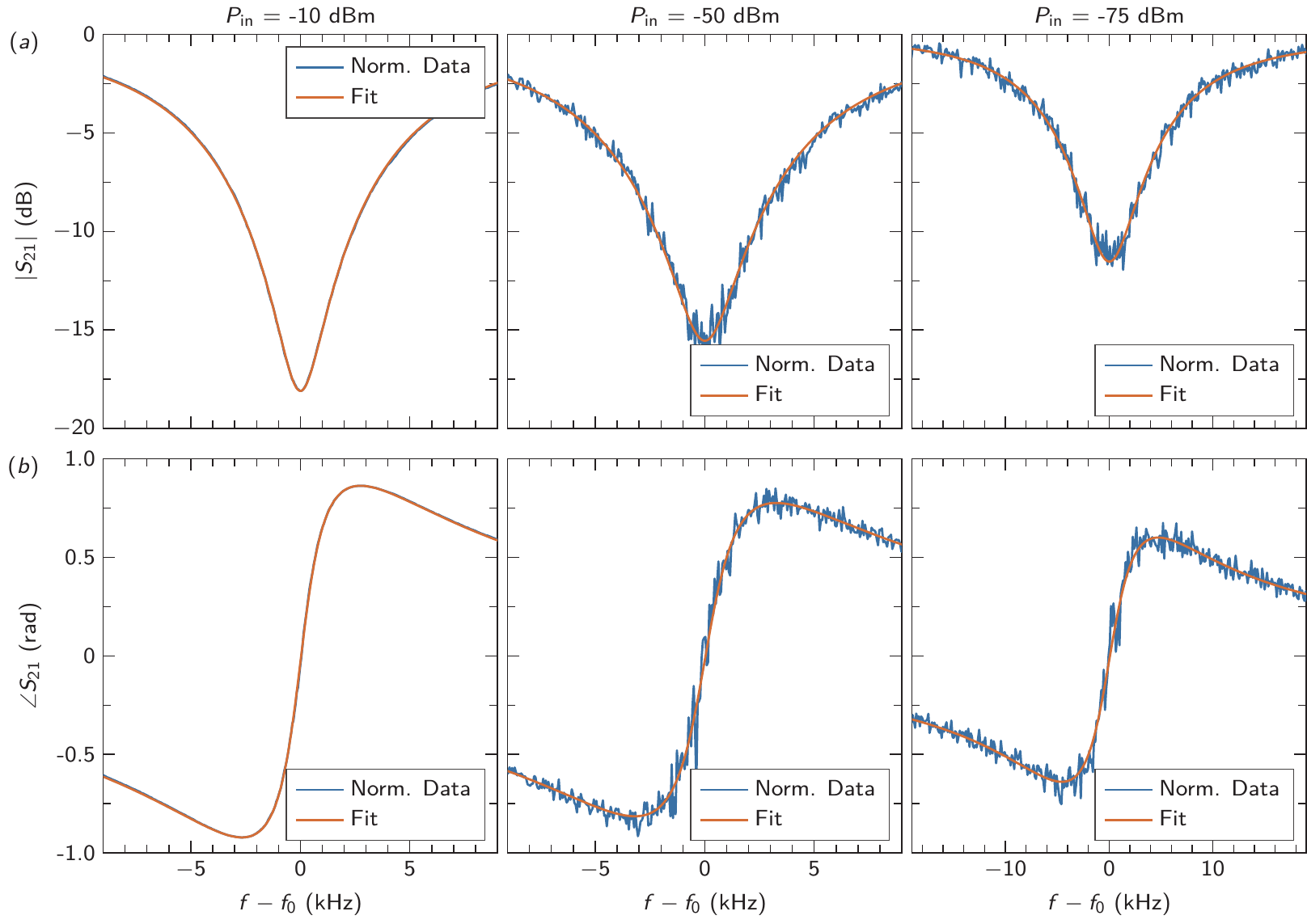}
\caption{Resonator fitting procedure. Each panel comprises three columns, one for each different excitation power (see appendix text for details). (a) $| S_{21} |$ vs.~$f - f_0$; normalized resonator data (``Norm. Data''; blue (dark gray)) and fitting curves (``Fit''; orange (light gray)). (b) Same as in~(a), but for~$\angle S_{21}$ vs.~$f - f_0$. The fitting curves are obtained by first inverting the normalized resonator data, fit as explained in the appendix text, re-invert the fitting data, and then plot them over the normalized resonator data.}
	\label{figure7a_7b_earnest}
\end{figure*}

The resonator fitting procedure takes as inputs transmission-coefficient (scattering parameter (or S-parameter)) $S_{21}$ data over two frequency ranges: One used for normalization; another used for resonator fitting. The \textit{normalization data} is measured away from resonance, in the flat transmission regions on the left and right side of the resonance dip. Figure~\ref{figure5_earnest} shows the two spots where the normalization data is taken for the transmission-coefficient magnitude~$| S_{21} |$, including a fitting line that connects them. The \textit{resonator data} is measured close to resonance, within the resonance dip (see figure~\ref{figure5_earnest}). We use the following fitting function~\cite{Megrant:2012}:
\begin{equation}
\tilde{S}_{21}^{-1} = 1 + \frac{Q_{\textrm{i}}}{Q^{\ast}_{\textrm{c}}} e^{i \phi} \frac{f_0}{f_0 + i 2 Q_{\textrm{i}} \Delta f} \, ,
	\label{eq:S:21:-1}
\end{equation}
where~$Q^{\ast}_{\textrm{c}}$ is the rescaled coupling quality factor of the resonator, $\phi$ is an offset angle, and $\rmi^2 = -1$.

As initial fitting parameters we use~$Q_{\textrm{i}} = \SI{e+06}{}$, $Q^{\ast}_{\textrm{c}} = \SI{3e+05}{}$, and $\phi = 0$. Additionally, we choose~$f_0$ to be the frequency where~$| S_{21} |$ reaches a minimum.

The resonator fitting procedure comprises the following steps:
\begin{enumerate}[(1)]
	\item Take the~$S_{21}$ normalization and resonator data as input data, as shown in figure~\ref{figure6a_6b_earnest};
    \item Fit the normalization data in magnitude and phase separately to a line, as shown also in figure~\ref{figure6a_6b_earnest};
	\item Extrapolate the value of the magnitude and phase in the frequency range of the resonator data;
	\item Subtract the extrapolated data from the resonator data in the complex plane. This has the effect of de-embedding the measurement setup from the data. The magnitude and phase of the normalized resonator data is shown in figure~\ref{figure7a_7b_earnest};
	\item Invert the normalized resonator data;
	\item Fit the inverted normalized resonator data to~(\ref{eq:S:21:-1}) using standard non-linear least squares techniques, starting from a reasonable initial parameters choice;
	\item Calculate the fitting errors from the covariance matrix;
	\item Obtain the fitting parameters and associated errors.
\end{enumerate}

\subsection{Resonator measurement procedure}
	\label{app:Resonator:measurement:procedure}

The resonator measurement procedure takes as inputs an approximate~$f_0$ and $\Delta f = f - f_0$, as well as a list of VNA excitation powers~$\{ P^{j}_{\textrm{in}} \}$ (where~$j = 1, 2, \ldots, N$ and $N$ is an integer number) and intermediate frequency~(IF) bandwidths~$\{ \Delta f^{j}_{\textrm{IF}} \}$. The procedure acquires a sufficient number of traces at each excitation power to satisfy a preset fitting error threshold. In our measurements, the upper bound for this threshold ($Q_{\textrm{i}}$ error) is set to~\SI{1}{\percent}. The threshold, the number of points to measure for the resonator data and relative frequency span, and the maximum number of measured traces~$N_{\textrm{tr}}$ are configurable, although they are usually not changed from experiment to experiment. Typically, we measure one single trace at the highest excitation power and \mbox{$\sim 30$} traces at the lowest powers.

The value of~$\langle n_{\textrm{ph}} \rangle$ is estimated from the room-temperature power at the input channel, $P_{\textrm{in}}$, and the knowledge of the total input-channel attenuation coefficient~$\alpha_{\textrm{att}}$. A total attenuation of~\mbox{$\sim \SI{76}{\decibel}$} corresponds to~$\alpha_{\textrm{att}} \approx \SI{3.98e+07}{}$. Following~\cite{Bejanin:2016}, we find
\begin{equation}
\langle n_{\textrm{ph}} \rangle = \frac{2}{h \pi^2} \frac{Q^2_{\ell}}{Q^{\ast}_{\textrm{c}}} \frac{P^{\prime}_{\textrm{in}}}{\tilde{f}^2_0} \, ,
	\label{eq:mean:n:ph}
\end{equation}
where~$h$ is the Planck constant,
\begin{equation}
\frac{1}{Q_{\ell}} = \frac{1}{Q_{\textrm{i}}} + \frac{1}{Q^{\ast}_{\textrm{c}}}
	\label{eq:1:Q:l}
\end{equation}
is the inverse loaded quality factor of the resonator, and $P^{\prime}_{\textrm{in}} = P_{\textrm{in}} / \alpha_{\textrm{att}}$ is the power at the resonator input. In our estimate, we use~$\alpha_{\textrm{att}} \approx \SI{3.98e+07}{}$ also for the dilution refrigerator at operating conditions.

The resonator measurement procedure comprises the following steps:
\begin{enumerate}[(1)]
	\item Set an approximate value of~$f_0$ and $\Delta f$ as input values;
	\item Acquire an~$S_{21}$ trace centered about the input value~$f_0$ and with span~$\Delta f$;
	\item Identify the resonator by fitting the data. If the fitting fails, abort; if it succeeds, update the value of~$f_0$ and choose the new frequency span to be~$10 \times f_0 / Q_{\textrm{i}}$;
	\item Perform a loop over the values~$\{ P^{j}_{\textrm{in}} \}$:
	\begin{enumerate}[(a)]
		\item Set the~$j$th value~$P^{j}_{\textrm{in}}$ and the corresponding~$\Delta f^{j}_{\textrm{IF}}$;
		\item Continue the loop as long as the fitting error is above the preset threshold or until~$N_{\textrm{tr}}$ is reached:
		\begin{enumerate}[(i)]
			\item Acquire a trace and average it with the previously acquired trace for the present value of~$P^{j}_{\textrm{in}}$;
			\item Fit the averaged traces;
		\end{enumerate}
		\item If the loop is completed successfully (i.e., the fitting error is below threshold), update~$f_0$ and $\Delta f$ as before and move to the next excitation power; otherwise abort the procedure.
	\end{enumerate}
\end{enumerate}

Figures~\ref{figure6a_6b_earnest} and \ref{figure7a_7b_earnest} show normalization and resonator data as well as fitting curves obtained from the inverse fitting procedure for three different excitation powers, $P_{\textrm{in}} = - 10, - 50, \textrm{ and } \SI{-75}{\decibel}$-milliwatts~(dBm), and corresponding IF bandwidths, $\Delta f_{\textrm{IF}} = 300, 100, \textrm{ and } \SI{5}{\hertz}$.

\vspace{5.0mm}

\section*{References}

%\bibliography{bibliography_earnest}
\bibliographystyle{iopart-num}

\providecommand{\newblock}{}

\end{document}